\documentclass[a4paper,12pt]{article}
\usepackage{epsfig}
\usepackage{rotating}
\usepackage{cite}

\begin{document}
\newcommand{\omclb}      {\ensuremath{\mathrm {1-CL_b}}}
\newcommand{\clb}        {\ensuremath{\mathrm {CL_b}}}
\newcommand{\cls}        {\ensuremath{\mathrm {CL_s}}}
\newcommand{\clsb}       {\ensuremath{\mathrm {CL_{s+b}}}}
\newcommand{\Z}{\ensuremath{\mathrm{Z}}}
\newcommand{\Zo}{\ensuremath{\mathrm{Z^0}}}
\newcommand{\W}{\ensuremath{\mathrm{W}}}
\newcommand{\bZo}{{\boldmath \mbox{$\mathrm{Z}$}}}
\newcommand{\Zg}{\ensuremath{\mathrm{Z}\gamma}}
\newcommand{\Zgs}{\ensuremath{(\mathrm{Z}/\gamma)^{*}}}
\newcommand{\ZZ}{\ensuremath{\mathrm{Z}\mathrm{Z}}}
\newcommand{\WW}{\ensuremath{\mathrm{WW}}}
\newcommand{\Zs}{\ensuremath{\mathrm{Z}^{*}}}
\newcommand{\hZ}{\ensuremath{\mathrm{hZ}}}
\newcommand{\h}{\ensuremath{\mathrm{h}}}
\newcommand{\Ho}{\ensuremath{\mathrm{H^{0}}}}
\newcommand{\ho}{\ensuremath{\mathrm{h^0}}}
\newcommand{\Hp}{\ensuremath{\mathrm{H}^{+}}}
\newcommand{\Hm}{\ensuremath{\mathrm{H}^{-}}}
\newcommand{\Hsm}{\ensuremath{\mathrm{H}^{0}_{\mathrm{SM}}}}
\newcommand{\A}{\ensuremath{\mathrm{A}^{0}}}
\newcommand{\Hpm}{\ensuremath{\mathrm{H}^{\pm}}}
\newcommand{\X}{\ensuremath{{\tilde{\chi}^0}}}
\newcommand{\ko}{\ensuremath{{\tilde{\chi}^0}}}
\newcommand{\ee}{\ensuremath{\mathrm{e}^{+}\mathrm{e}^{-}}}
\newcommand{\mumu}{\ensuremath{\mu^{+}\mu^{-}}}
\newcommand{\tautau}{\ensuremath{\tau^{+}\tau^{-}}}
\newcommand{\nn}{\ensuremath{\nu \bar{\nu}}}
\newcommand{\qq}{\ensuremath{\mathrm{q} \bar{\mathrm{q}}}}
\newcommand{\glgl}{\ensuremath{\mathrm{gg}}}
\newcommand{\invpb}{\ensuremath{\mathrm{pb}^{-1}}}
\newcommand{\ra}{\ensuremath{\rightarrow}}
\newcommand{\br}{\ensuremath{\boldmath {\rightarrow}}}
\newcommand{\ga}{\ensuremath{\gamma}}
\newcommand{\gamgam}{\ensuremath{\gamma\gamma}}
\newcommand{\tp}{\ensuremath{\tau^+}}
\newcommand{\tm}{\ensuremath{\tau^-}}
\newcommand{\tpm}{\ensuremath{\tau^{\pm}}}
\newcommand{\uu}{\ensuremath{\mathrm{u} \bar{\mathrm{u}}}}
\newcommand{\dd}{\ensuremath{\mathrm{d} \bar{\mathrm{d}}}}
\newcommand{\ssbar}{\ensuremath{\mathrm{s} \bar{\mathrm{s}}}}
\newcommand{\bb}{\ensuremath{\mathrm{b} \bar{\mathrm{b}}}}
\newcommand{\cc}{\ensuremath{\mathrm{c} \bar{\mathrm{c}}}}
\newcommand{\csbar}{\ensuremath{\mathrm{c} \bar{\mathrm{s}}}}
\newcommand{\cbars}{\ensuremath{\bar{\mathrm{c}}\mathrm{s}}}
\newcommand{\nubar}{\ensuremath{\bar{\nu}}}
\newcommand{\mQ}{\ensuremath{m_{\mathrm{Q}}}}
\newcommand{\mZ}{\ensuremath{m_{\mathrm{Z}}}}
\newcommand{\mH}{\ensuremath{m_{\mathrm{H}}}}
\newcommand{\mHrec}{\ensuremath{m_{\mathrm{H}}^{\mathrm{rec}}}}
\newcommand{\mHp}{\ensuremath{m_{\mathrm{H}^+}}}
\newcommand{\mh}{\ensuremath{m_{\mathrm{h}}}}
\newcommand{\mA}{\ensuremath{m_{\mathrm{A}}}}
\newcommand{\mHpm}{\ensuremath{m_{\mathrm{H}^{\pm}}}}
\newcommand{\mHsm}{\ensuremath{m_{\mathrm{H}^0_{\mathrm{SM}}}}}
\newcommand{\mW}{\ensuremath{m_{\mathrm{W}^{\pm}}}}
\newcommand{\mt}{\ensuremath{m_{\mathrm{t}}}}
\newcommand{\mb}{\ensuremath{m_{\mathrm{b}}}}
\newcommand{\lpm}{\ensuremath{\ell ^{+} \ell^{-}}}
\newcommand{\GeV}{\ensuremath{\ \mathrm{GeV}}}
\newcommand{\Gc}{\ensuremath{\ \mathrm{GeV}/\mathrm{c}}}
\newcommand{\Gcs}{\ensuremath{\ \mathrm{GeV}/\mathrm{c}^2}}
\newcommand{\Mcs}{\ensuremath{\ \mathrm{MeV}/\mathrm{c}^2}}
\newcommand{\mm}{\ensuremath{\ \mathrm{mm}}}
\newcommand{\sba}{\ensuremath{\sin ^2 (\beta -\alpha)}}
\newcommand{\cba}{\ensuremath{\cos ^2 (\beta -\alpha)}}
\newcommand{\tanb}{\ensuremath{\tan \beta}}
\newcommand{\sqrts}{\ensuremath{\sqrt {s}}}
\newcommand{\sqrtsp}{\ensuremath{\sqrt {s'}}}
\newcommand{\msusy}{\ensuremath{M_{\mathrm{SUSY}}}}
\newcommand{\mg}{\ensuremath{$m_{\tilde{\mathrm{g}}}$}}

\begin{titlepage}
\centerline{\Large EUROPEAN ORGANIZATION FOR NUCLEAR RESEARCH}

\begin{flushright}
       CERN-EP/2003-081   \\ 28 November 2003
\end{flushright}

\bigskip\bigskip\bigskip\bigskip\bigskip
\begin{boldmath}
\begin{center}{\Large \bf Flavour independent search for Higgs bosons decaying
into hadronic final states in \ee\ collisions at LEP}
\end{center}
\end{boldmath}
\bigskip

\vspace{0.1cm}  
\begin{center}
 {\Large {The OPAL Collaboration}} \\
\end{center}

\bigskip
\begin{center}{\large  Abstract}\end{center}

A search for the Higgsstrahlung process $\ee\ra\hZ$ is described, where the 
neutral Higgs boson h is assumed to decay into hadronic final states. In order to be
sensitive to a broad range of models, the search is performed 
independent of the flavour content of the Higgs boson decay. The analysis
is based on \ee\ collision data collected by the OPAL detector at energies
between $192\GeV$ and $209\GeV$. The search does not reveal any significant excess over
the Standard Model background prediction. Results are combined with previous
searches at energies around $91\GeV$ and at $189\GeV$.
A limit is set on the product of the
cross-section and the hadronic branching ratio of the Higgs boson, as a function
of the Higgs boson mass. Assuming the hZ coupling predicted 
by the Standard Model, and a Higgs boson decaying
only into hadronic final states, a lower bound of $104\Gcs$ is set on the
mass at the $95\%$ confidence level. 

\bigskip\bigskip\bigskip\bigskip\bigskip\bigskip\bigskip\bigskip
\begin{center}
  {\LARGE
   (Submitted to Physics Letters {\bf B})}
\end{center}
\end{titlepage}

\newpage

\begin{center}{\Large        The OPAL Collaboration
}\end{center}\bigskip
\begin{center}{
G.\thinspace Abbiendi$^{  2}$,
C.\thinspace Ainsley$^{  5}$,
P.F.\thinspace {\AA}kesson$^{  3,  y}$,
G.\thinspace Alexander$^{ 22}$,
J.\thinspace Allison$^{ 16}$,
P.\thinspace Amaral$^{  9}$, 
G.\thinspace Anagnostou$^{  1}$,
K.J.\thinspace Anderson$^{  9}$,
S.\thinspace Arcelli$^{  2}$,
S.\thinspace Asai$^{ 23}$
% ...
}\end{center}\bigskip
\bigskip
$^{  1}$School of Physics and Astronomy, University of Birmingham,
Birmingham B15 2TT, UK
\newline
$^{  2}$Dipartimento di Fisica dell' Universit\`a di Bologna and INFN,
I-40126 Bologna, Italy
\newline
$^{  3}$Physikalisches Institut, Universit\"at Bonn,
D-53115 Bonn, Germany
\newline
% ...
$^{  5}$Cavendish Laboratory, Cambridge CB3 0HE, UK
\newline
% ...
$^{  9}$Enrico Fermi Institute and Department of Physics,
University of Chicago, Chicago IL 60637, USA
\newline
% ...
$^{ 16}$Department of Physics, Schuster Laboratory, The University,
Manchester M13 9PL, UK
\newline
% ...
$^{ 22}$Department of Physics and Astronomy, Tel Aviv University,
Tel Aviv 69978, Israel
\newline
$^{ 23}$International Centre for Elementary Particle Physics and
Department of Physics, University of Tokyo, Tokyo 113-0033, and
Kobe University, Kobe 657-8501, Japan
\newline
% ...
$^{  c}$ and Institute of Nuclear Research, Debrecen, Hungary
\newline
$^{  e}$ and Department of Experimental Physics, University of Debrecen, 
Hungary
\newline
% ...
$^{  g}$ and Research Institute for Particle and Nuclear Physics,
Budapest, Hungary
\newline
$^{  h}$ now at University of Liverpool, Dept of Physics,
Liverpool L69 3BX, U.K.
\newline
% ...
$^{  q}$ now at IPHE Universit\'e de Lausanne, CH-1015 Lausanne, Switzerland
\newline
% ...
$^{  s}$ now at Universitaire Instelling Antwerpen, Physics Department, 
B-2610 Antwerpen, Belgium
\newline
$^{  t}$ now at RWTH Aachen, Germany
\newline
$^{  u}$ and High Energy Accelerator Research Organisation (KEK), Tsukuba,
Ibaraki, Japan
\newline
% ...
$^{  y}$ now at CERN
\newline
% ...

\section{Introduction}
In the Standard Model (SM) and for masses relevant to the LEP
energy range, the Higgs boson is predicted to be produced principally by the
Higgsstrahlung process $\ee\ra\hZ$ and to decay dominantly into the \bb\ 
channel. This is also the case in large domains of the Minimal Supersymmetric Standard Model
(MSSM) parameter space (the Higgs phenomenology is reviewed, e.g. in 
Ref.~\cite{hunter}). Most of the searches conducted so far at LEP,
therefore, tag the b flavour to enhance 
the Higgs boson signal. 

In other scenarios, however, the decay of the Higgs boson into lighter quark
flavours or into gluon pairs may be important. Such is the case in general models
with two Higgs field doublets (2HDM)~\cite{hunter,Hollik} or other extended
models~\cite{composite}. In order to be sensitive to Higgs bosons predicted by such
models, the search described here is based only on kinematic selections which are 
insensitive to the hadron flavour present in the final state. 
Such searches have already been reported by OPAL; these were based on data
collected at energies close to the Z boson resonance~\cite{opalSMHiggsLEP1} and at a
centre-of-mass energy (\sqrts) of $189\GeV$~\cite{opal2HDM189}. A similar search has
also been reported by ALEPH~\cite{ALEPHFlavIndep}.

This paper describes a flavour independent search which is based on OPAL data collected at
centre-of-mass energies between $192$~and $209\GeV$ with an integrated 
luminosity of about $420\invpb$. For the results presented,
this search is combined with the earlier OPAL searches~\cite{opalSMHiggsLEP1,opal2HDM189}.
\section{Data sets and Monte Carlo simulation}
The OPAL detector is described in Ref.~\cite{detector}. The events selected for
the analysis have to satisfy a set of detector status requirements which ensure
that all relevant detector elements are active.
Events are reconstructed from charged particle tracks observed in the central
tracking detector and energy deposits
({}``clusters'') in the electromagnetic and hadron calorimeters.
The tracks and clusters are required to pass a set of quality requirements 
~\cite{higgsold}. In calculating the visible energies and momenta
$E_{\mathrm{vis}}$
and $\vec{P}_{\mathrm{vis}}$, either for individual jets or for the events, 
corrections
are applied to prevent double-counting of the energy attributed to the 
tracks and to the clusters geometrically associated to the tracks~\cite{lep2neutralino}.

The data sets to which the present analysis applies were collected in $1999$
at \sqrts\ between $192$ and $202\GeV$ and in the year $2000$ at \sqrts\ between $200$
and $209\GeV$. After the detector status requirements the data sample 
has an integrated luminosity of approximately $420\invpb$. The exact amount varies
among the different channels (see Table\ \ref{tab:flowSum}).

A variety of Monte Carlo samples have been generated to estimate the selection
efficiencies for the Higgs boson signal and for the background processes.
In order to cover the range of energies of the data, the simulations are
performed at fixed values of \sqrts\ between $192$ and $210\GeV$ and for a set of
Higgs boson masses. Spline fits are used 
to calculate the signal efficiencies at intermediate values. 

The Higgsstrahlung process is modelled with the HZHA generator~\cite{hzha}.
Samples of 1000 to 5000 events were produced at fixed masses, between $30$
and $120\Gcs$. The Higgs boson is required to decay, either according to the SM, 
or separately to \cc, \ssbar or to pairs of  gluons.

The simulated background samples typically have more than $30$ times the 
statistics of the collected data. The following event generators are used:
KK2F~\cite{mcKK2F}
and PYTHIA~\cite{pythia} for the process $\qq(\gamma)$, 
grc4f~\cite{grc4f}, KORALW~\cite{koralw}
and EXCALIBUR~\cite{excalibur} for the four-fermion processes, 
BHWIDE~\cite{bhwide} for $\ee(\gamma)$, KORALZ~\cite{koralz} for 
$\mumu(\gamma)$ and $\tautau(\gamma)$,
and PHOJET~\cite{phojet}, HERWIG~\cite{mcHERWIG} and VERMASEREN~\cite{vermaseren}
for hadronic and leptonic two-photon processes and for $\ee\ra\ee\gamma\gamma$.
Hadronisation is modelled with JETSET~\cite{pythia} using parameters described in~\cite{opaltune}. The cluster fragmentation
model in HERWIG is used to study the uncertainties due to quark and
gluon jet fragmentation. The Monte Carlo samples pass through a detailed
simulation of the OPAL detector~\cite{gopal} and are subjected to the same
analysis procedure as applied to the data.
\section{Analysis}
The search described in this paper addresses the Higgsstrahlung process
$\ee\ra\hZ$. The neutral Higgs boson h is assumed to decay into
quark pairs of arbitrary flavour or into gluon pairs. The following $\hZ$ final states
(search channels) are therefore considered, depending on the decay of the $\Z$ boson: 
the four-jet channel ($\Z\ra\qq$), 
the missing energy channel ($\Z\ra\nn$) and the electron, 
muon and tau channels ($Z\ra\ee,\ \mumu$\ and \tautau). 

The analysis assumes that the decay width of the Higgs boson is within the
range $10^{-4} < \Gamma_\h < 1\Gcs$. This ensures that the decay of the Higgs
boson occurs within about $1\mm$ of the \ee\ interaction point and that the 
reconstructed Higgs boson mass has a width that is dominated by the
experimental resolution, between $2$ and $5\Gcs$, depending on the search channel.
The search strategies are similar
to those applied by OPAL in the search for the SM Higgs boson~\cite{sm-final} 
(see Ref.~\cite{opal183} for the missing energy channel) except
that the b-tagging requirements are replaced by more elaborate kinematic
selections.

In the searches addressing each of the final states,
a preselection is applied first which strongly reduces
the background while maintaining a high signal detection efficiency.
The preselected events are then submitted to a likelihood
test, which discriminates between the signal and the 
two most important background sources, 2-fermion and 4-fermion processes.
Other background processes, in particular 2-photon events, are negligible
after the first preselection cuts (see Section 3 of~\cite{sm-final}). 
The likelihood function is constructed 
from reference distributions of a number of discriminating variables 
which are obtained from detailed simulations of the signal and background 
processes. In the four-jet channel, these distributions are
obtained from a three-dimensional spline fit to the distributions of simulated events where the 
dimensions are \sqrts, the hypothetic Higgs boson mass (test-mass) and the variable itself. 

Finally, a cut is applied on the value of this likelihood function.
For each of the search channels, the effect of the preselection and likelihood 
cuts on the data samples, the total background and its contributions, 
and on the signal detection efficiency for two
test-masses can be followed through Table~\ref{tab:flowSum}.  
\begin{table}
\begin{center}{
    \begin{tabular}{|l|r|r|r|r|cc|}
      \hline 
      Cut &
      Data &
      Total &
      $\qq(\gamma)$ &
      4--fermi. &
      \multicolumn{2}{|c|}{Efficiency in \%}\\
      &
      &
      bkg.&
      &
      &
      $90\Gcs$ & $100\Gcs$\\
      \hline
      \hline
      \multicolumn{7}{|c|}{Four--jet Channel ~~ Luminosity = $424.3\invpb$}\\
      \hline
(1) & 39090 & 38831.1 & 29929.3 & 8322.0  & 100 & 100 \\
(2) & 13692 & 13648.5 & 8602.5 & 5012.2 & 100 & 100 \\
(3) & 4645  &  4504.3 & 1077.9 & 3418.4  & 93  & 95  \\
(4) & 4200  &  4038.4 &  932.7 & 3105.7  & 92  & 94  \\
(5) & 3695  &  3561.3 &  603.2 & 2958.1  & 90  & 91  \\
(6) & 3594  &  3447.2 &  581.2 & 2866.0  & 89  & 90  \\
(7) & 2535  &  2399.6 &  504.2 & 1895.4  & 81  & 83  \\
(8) & 2081  &  1975.3 &  477.2 & 1498.1  & 78  & 80  \\
(9) &  659  &   637.4 &  155.8 &  481.6  & 59  & 66  \\
\hline
${\cal L}$ & 439 & 414.0 & 103.8 & 136.0  & 52  & 54 \\
      \hline\hline
      \multicolumn{7}{|c|}{Missing--energy Channel ~~ Luminosity = $420.9\invpb$}\\
      \hline
(1) & 9040 & 8524.6 & 6063.7 & 2382.4 & 87 & 78 \\
(2) & 2615 & 2391.3 &  686.0 & 1691.2 & 80 & 73 \\
(3) & 2462 & 2289.9 &  665.4 & 1614.6 & 77 & 73 \\
(4) & 1635 & 1598.4 &  110.7 & 1487.7 & 72 & 69 \\
(5) &  650 &  605.4 &   48.5 &  556.8 & 70 & 67 \\
(6) &  298 &  291.4 &   42.3 &  249.1 & 65 & 62 \\
\hline
${\cal L}$ & 123 & 133.1 & 6.3 & 126.6 & 45 & 48 \\
      \hline\hline
      \multicolumn{7}{|c|}{Electron Channel ~~ Luminosity = $422.3\invpb$}\\
      \hline
(1) & 18042 & 18221.3 & 12176.4 & 6045.0 & 92 & 97 \\
(2) &   558 &   538.7 &   252.8 &  286.1 & 75 & 78 \\
(3) &   429 &   378.6 &   171.0 &  207.6 & 74 & 78 \\
\hline
${\cal L}$ & 23 & 16.6 &  0.2   &   16.3 & 59 & 59 \\
      \hline\hline
      \multicolumn{7}{|c|}{Muon Channel ~~ Luminosity = $421.4\invpb$}\\
      \hline
(1) & 18008 & 18184.6 & 8715.5 & 9469.0 & 88 & 92 \\
(2) &   505 &   477.5 &  236.5 &  241.0 & 77 & 81 \\
(3) &    79 &    66.1 &   32.6 &   33.6 & 74 & 75 \\
\hline
${\cal L}$ & 16 & 15 & 6.6 & 8.4 & 64.8 & 62.4 \\
      \hline\hline
      \multicolumn{7}{|c|}{Tau Channel ~~ Luminosity = $409.0\invpb$}\\
      \hline
(1) & 10417 & 10082 & 5520.1 & 4561.8 & 83 & 78 \\
(2) &  1652 & 1687.6 & 187.0 & 1500.9 & 62 & 61 \\
(3) &   418 &  404.5 &  99.5 &  305.2 & 48 & 47 \\
(4) &   358 &  343.1 &  96.6 &  246.3 & 47 & 47 \\
\hline
${\cal L}$ & 3 & 8.8 & 0.23 & 8.57 & 27 & 21 \\
      \hline
    \end{tabular}} 
\end{center}

      \vspace{-0.5cm}

\caption{\label{tab:flowSum} 
      Number of events selected in the different 
      search channels after consecutive cuts. In each case, the final
      likelihood cut is denoted by $\mathcal{L}$.
      The number of events found in the data is compared
      to the expectation from simulations. 
      In the four-jet channel the
      numbers up to and including cut\ (8) are valid for all test-masses; 
      those for cut (9) and the final likelihood cut 
      are given for a test-mass of $100\Gcs$.
      The last two columns show the evolution of the selection efficiencies 
      for Higgs bosons of $90\Gcs$\ and $100\Gcs$\ mass decaying exclusively 
      into hadronic final states at $196$ and $206\GeV$ centre-of-mass energy, 
      respectively.}
\end{table}

\begin{figure}[htb]
\begin{center}
\epsfig{figure=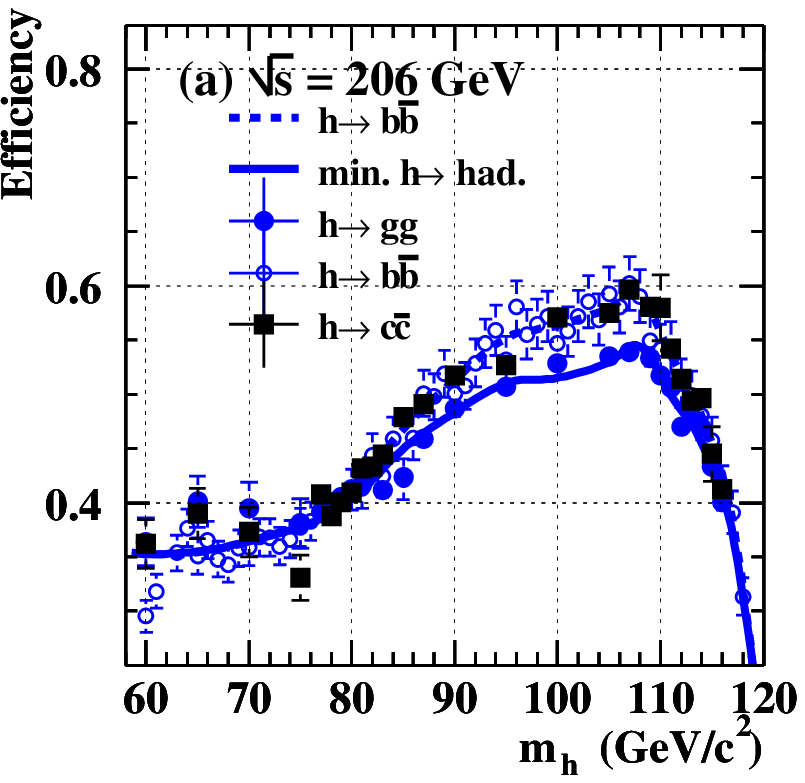,width=0.49\textwidth}
\epsfig{figure=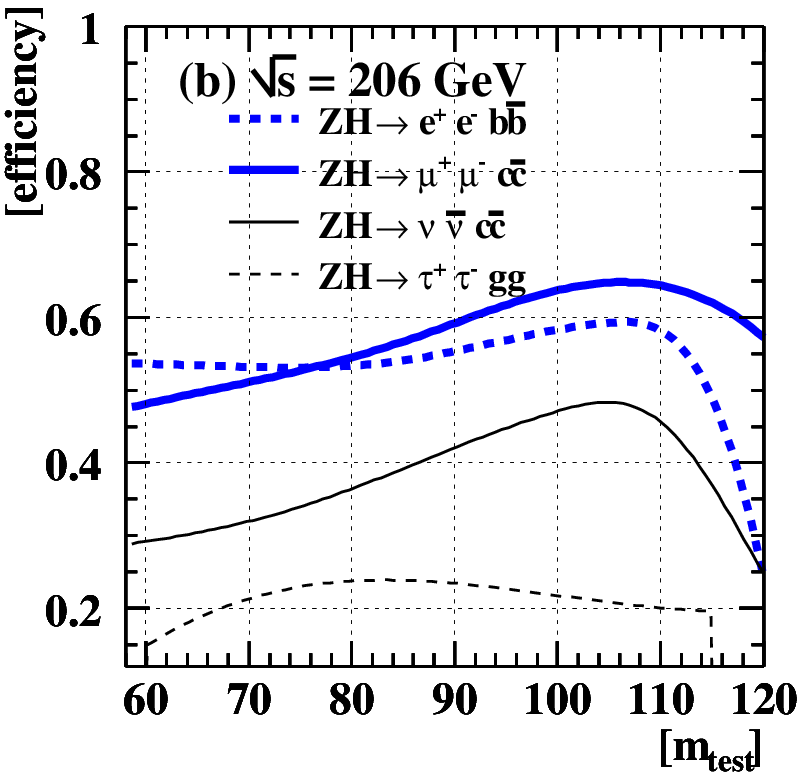,width=0.49\textwidth}
\caption[]{\small  Selection efficiencies for the Higgsstrahlung process
in the different search channels, at $\sqrts=206\GeV$. 
(a) Four-jet channel, flavour-dependence. The full line shows the result from a
spline fit to the points with the lowest efficiency. (b) All but the four-jet channel.
In each case, the lowest of the efficiencies over all hadron flavours 
is plotted.}
\label{figure1}
\end{center}
\end{figure}

The signal efficiency is evaluated separately for each of the $\h\ra\bb$,
\cc, \ssbar\ and \glgl\ decay hypotheses. At a given test-mass, these efficiencies 
typically vary by about $\pm 5\%$. This is illustrated in
Figure~\ref{figure1}(a) for the search in the four-jet channel.
In deriving flavour independent bounds on the $\hZ$ coupling, 
the smallest of these efficiencies is used; it is obtained for h\ra~gg in the 
four-jet and tau channels and for $\h\ra\cc$\ or \bb\ in the missing energy 
and lepton channels.
For the decays into light flavours, $\h\ra\uu$\ and \dd, the efficiencies 
are slightly higher since the jets are better collimated 
and because weak semileptonic decays are absent;
this has been verified explicitly  using $\ee\ra\ZZ$ events.
These minimal efficiencies are shown for all but the four-jet 
search by the curves in Figure~\ref{figure1}(b).
\subsection{Search in the four-jet channel}
\begin{figure}[htb]
\begin{center}
\epsfig{figure=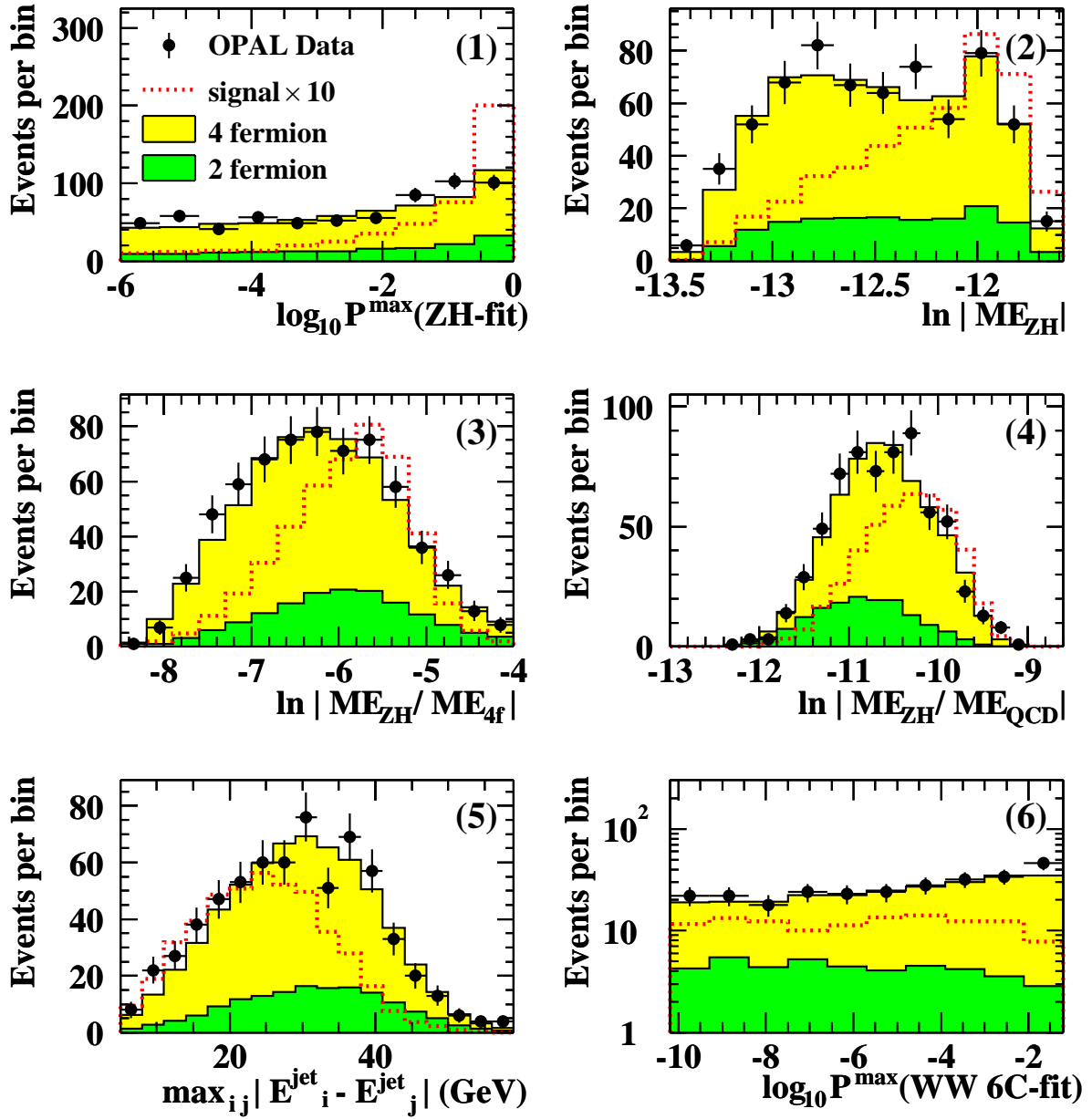,width=\textwidth}
\caption[]{\small Distributions of discriminating variables
which have been used in the construction of the signal
likelihood in the four-jet channel, the test-mass \mh\ being fixed at $100\Gcs$. The dots with error bars show the data.
The light and dark shaded histograms show
the expected background from four- and two-fermion processes.
The dashed histograms show the signal, scaled by a factor ten,
expected for a Higgs boson of $100\Gcs$ mass, with $\hZ$ coupling predicted by
the SM and decaying only into hadronic final states. }
\label{figure2}
\end{center}
\end{figure}

\begin{figure}[htb]
\begin{center}
\epsfig{figure=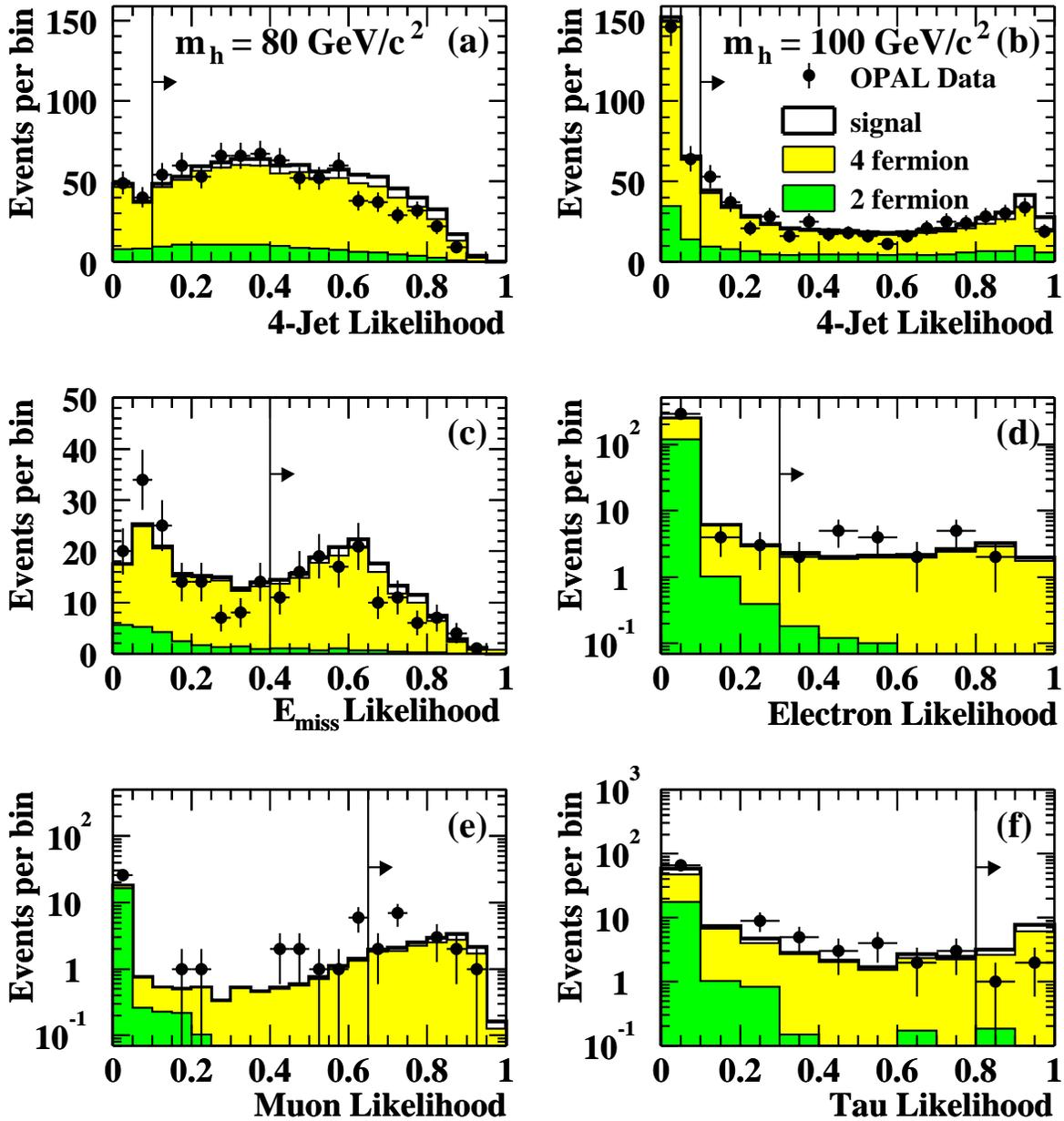,width=\textwidth}
\caption[]{\small Distributions of the signal 
likelihoods for the searches in the
(a-b) four-jet channel, (c) missing energy, (d) electron, (e) muon and 
(f) tau channels. In part (a) the test-mass \mh\ is fixed to $80\Gcs$; in all
other parts it is at $100\Gcs$. 
The points with error bars represent the data. 
The light and dark shaded histograms show
the expected background from four- and two-fermion processes.
The white histograms added on top of the background contributions
show the signal
expected for a Higgs boson of $100\Gcs$ mass ($80\Gcs$ in part (a)), 
with hZ coupling predicted by
the SM and decaying only into hadronic final states.
In each case, the vertical line indicates the final likelihood cut. }
\label{figure3}
\end{center}
\end{figure}

In the four-jet channel the main background arises from the $\ee\ra\WW$ process. 
Further contributions are from $\ee\ra\Zgs\ra\qq$ and 
$\ee\ra\ZZ$. The analysis described below is repeated for fixed 
test-masses, in steps of $250\Mcs$, between $60$ and $120\Gcs$.
The following preselection is applied:
\begin{enumerate}
\item Events must be identified as multihadronic final states \cite{hadronic}.
\item The effective centre-of-mass energy $\sqrtsp$ (disregarding initial-state
photon radiation, see Ref.~\cite{hadronic}),
is required to exceed $80\%$ of the total centre-of-mass energy.
\item Events are forced into four jets using the Durham
algorithm~\cite{jetDurham1} and are selected if the resolution parameter $y_{34}$ is
larger than $3\cdot 10^{-3}$.
\item Each of the jets must contain at least two charged particle tracks 
to suppress events with isolated leptons or photons, like $\ee\ra\qq\lpm$.
\item The matrix element $\mathrm{ME}_{\mathrm{QCD}}$ for the QCD-induced
processes $\ee\ra\qq\qq$\ and $\ee\ra\qq\glgl$ is 
calculated~\cite{QQQQEventWeight}, approximating the parton momenta 
by the reconstructed jet momenta. The matrix element
averaged over all possible flavour combinations is required to be within the
range $-3<\ln|\mathrm{ME}_{\mathrm{QCD}}|<-1$.
\item The $\chi^{2}$-probability resulting from a four-constraint (4C) 
kinematic fit which 
imposes energy and momentum conservation is required to be larger than $10^{-6}$.
\item The four-fermion background is reduced by a cut on the 
matrix element $\mathrm{ME}_{4\mathrm{f}}$ of the process $\ee\ra\qq\qq$,
calculated using EXCALIBUR~\cite{excalibur}. In the calculation the parton 
momenta are approximated by the jet momenta resulting from the 4C fit and
the matrix element is averaged over all flavour combinations. 
Its value is required to be within the range
$-8.5<\ln|\mathrm{ME}_{4\mathrm{f}}|<-4.9$.
\item The $\WW\ra\mathrm{hadrons}$ hypothesis is tested in a kinematic 6C fit 
imposing energy and momentum conservation and where the invariant masses 
of the two jet pairs are constrained to the W boson
mass. To suppress the WW background, the largest of the 
$\chi^{2}$-probabilities,
$\mathrm{P}^{\mathrm{max}}(\mathrm{WW})$,
for the three possible jet pairings is required to be less than
$6.3\%$.
\item Finally, for each value of the test-mass, a kinematic fit is performed
imposing energy and momentum conservation and constraining
one dijet mass to the test-mass and the other to the $\Z$ boson mass. 
In the fit, the reconstructed $\Z$ boson mass is allowed to vary within its
natural width according to a Gaussian distribution%
\footnote{The sensitivity of the search would be slightly lower 
if a Breit-Wigner distribution were used.}.
The largest of the $\chi^{2}$-probabilities $\mathrm{P}^{\mathrm{max}}(\mathrm{Zh})$
resulting from the six possible jet assignments to the $\Z$ and the $\h$ bosons 
is required to exceed $10^{-6}$.
\end{enumerate}
The signal likelihood is
constructed using the following 6 variables: (1)~the
maximum probability $\mathrm{P}^{\mathrm{max}}(\hZ)$ of the
$\hZ$ kinematic fit; (2)~the Higgsstrahlung matrix element $\mathrm{ME}_{\hZ}$
\cite{ZHMatrixElementTH} for the test-mass considered and for the jet 
combination which yields $\mathrm{P}^{\mathrm{max}}(\hZ)$;
the ratios (3)~$\mathrm{ME}_{\hZ}/\mathrm{ME}_{4\mathrm{f}}$
and (4)~$\mathrm{ME}_{\hZ}/\mathrm{ME}_{\mathrm{QCD}}$;
(5)~the difference between the maximum and minimum energies of the
four jets after the 4C kinematic fit; 
and (6) $\mathrm{P}^{\mathrm{max}}(\WW)$.
Distributions of these input variables are presented in Figure~\ref{figure2}, while
the likelihood distributions for two test-masses are shown in 
Figure~\ref{figure3} (a) and (b).
Events with a likelihood larger than $0.1$ are accepted. 
 
The signal efficiency and residual background rates are affected by the
following systematic uncertainties:
(a) Uncertainties in modelling of the momenta,
the angular and energy resolutions and the energy scale
of the reconstructed jets are
less than $2\%$ for both the signal efficiency and the background rate.
They have been determined by comparing calibration data taken at 
the Z resonance to the Monte Carlo simulation and transferring the observed
differences to the simulation of the high energy data.
(b) Uncertainties in modelling the preselection and likelihood variables are 
less than 3\% for the signal 
and $4-9\%$ for the background, depending on the test-mass.
Weights were applied to the simulated events such that
a $\chi^2<1$ is obtained when comparing the shapes of the distributions 
from the data and the simulation of the background (for each variable 
separately). The difference of the signal efficiency and background of
the weighted and unweighted events is considered as the systematic error.
It has been explicitly checked that a hypothetical signal is not hidden 
by this procedure.
(c) Using alternatively JETSET and HERWIG to simulate hadron
fragmentation yields a difference of $2-13\%$ for the background. 
(d) The cross-section of the four-fermion processes,
which dominates the residual background, is known to within $2\%$
\cite{FourFermionXSec}. 
(e) Monte Carlo statistics contribute $1-5\%$ for the signal and less than $3\%$
for the background. Combining all these
effects, the total systematic
uncertainty amounts to less than $6\%$ for the signal efficiency and $5-16\%$
for the residual background.

\begin{figure}[htb]
\begin{center}
\epsfig{figure=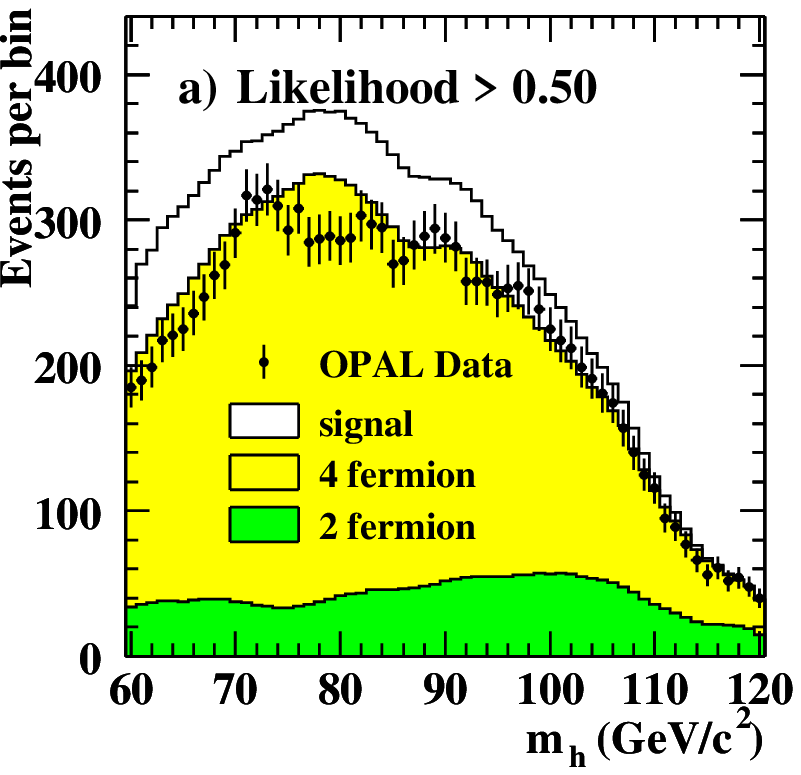,width=0.49\textwidth}
\epsfig{figure=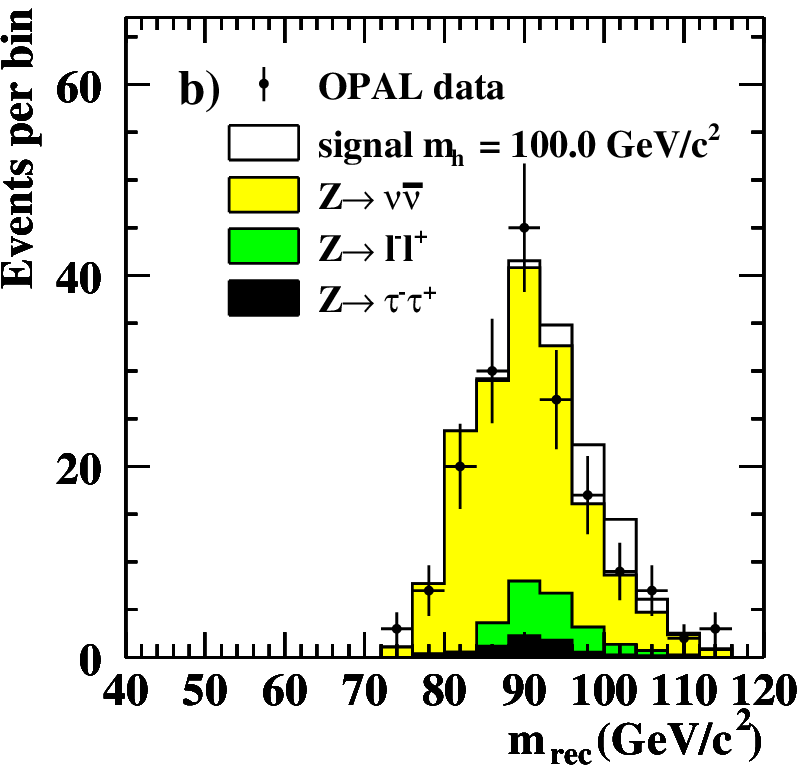,width=0.49\textwidth}\\
\caption[]{\small (a) Number of candidates selected in the
four-jet channel as a function of the test-mass \mh, together
with the predicted backgrounds and the signal 
from Higgsstrahlung added on top of the background. 
For the purpose of this figure 
the likelihood cut is raised to $0.5$.
(b) Combined distributions of the reconstructed Higgs boson mass
in the missing energy, electron, muon and tau channels. 
For the signal, the Higgs boson mass is fixed at $100\Gcs$.
In both parts of the figure, the hZ coupling 
predicted by the SM and $100\%$ hadronic Higgs boson
decays are assumed.}
\label{figure4}
\end{center}
\end{figure}

The number of selected events in the four-jet channel with a likelihood value
larger than 0.5 is shown in Figure~\ref{figure4} 
(a) for test-masses between $60$ and 
$120\Gcs$.
The selected data samples for mass hypotheses which differ by less than
the mass resolution (of about $5\Gcs$\ at high likelihood values)
are strongly correlated. For a test-mass of $100\Gcs$,
439 candidates pass the final likelihood cut of $0.1$ while $414\pm 53$
events are expected from background processes and $40$
events would be expected from Higgsstrahlung if the $\hZ$ coupling
predicted by the SM is assumed and the Higgs boson 
decays only into hadronic final states. The signal to background ratio
becomes more favourable for larger likelihood values.
\subsection{Search in the missing energy channel}
\begin{figure}[htb]
\begin{center}
\epsfig{figure=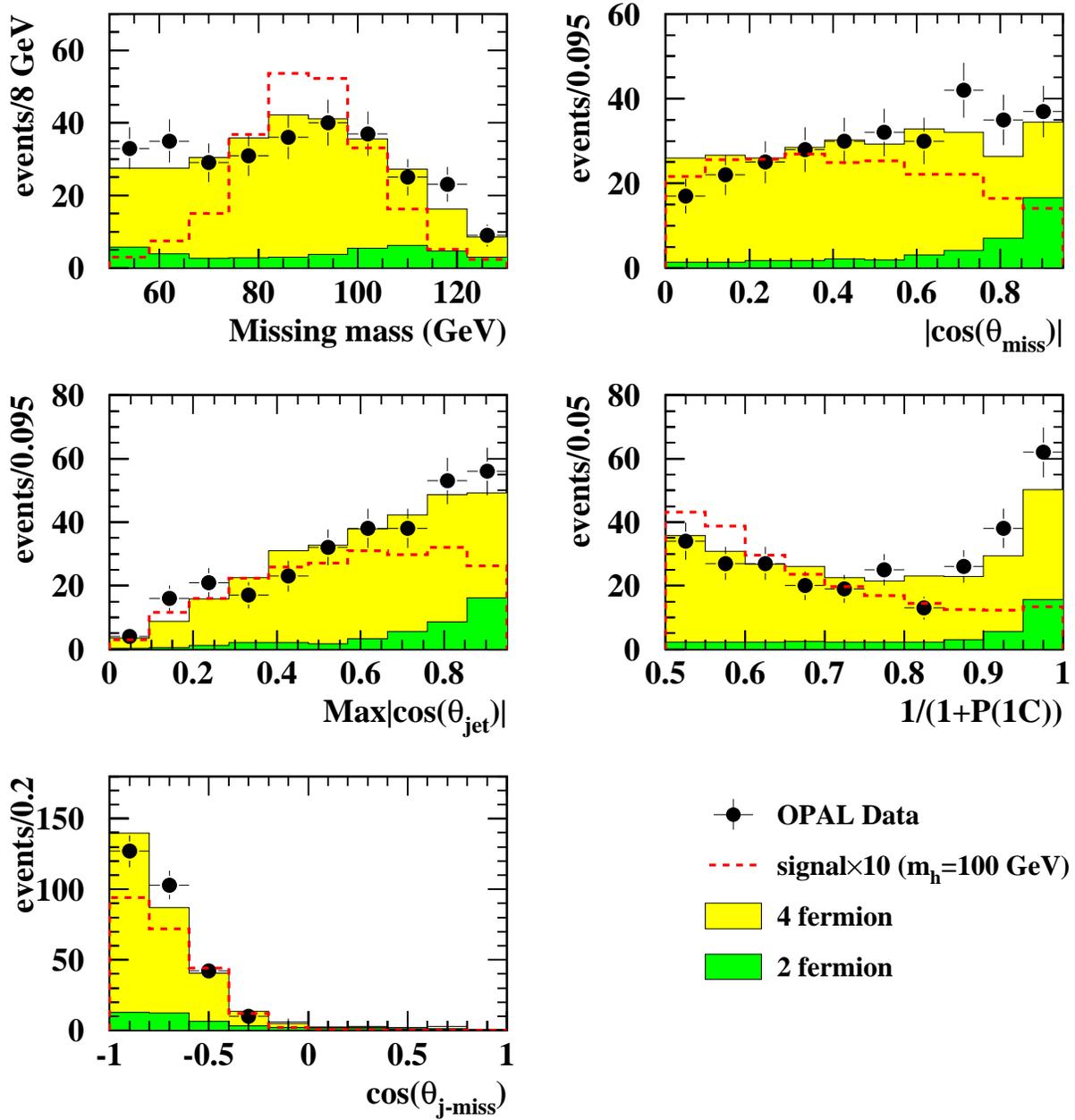,width=\textwidth}
\caption[]{\small Distributions of the discriminating 
variables used to calculate the signal likelihood
in the missing energy channel. 
The light and dark shaded histograms show
the expected background from four- and two-fermion processes.
The dashed histograms show the signal, scaled by a factor ten,
expected for a Higgs boson of $100\Gcs$ mass, with $\hZ$ coupling 
predicted by the SM and decaying only into hadronic final states.}
\label{figure5}
\end{center}
\end{figure}

Signal events in the missing energy channel are characterised by two
hadronic jets and a missing mass consistent with the Z boson
mass. The dominant backgrounds are four-fermion processes, in particular
from the semi-leptonic decays $\ee\ra\WW\ra\qq\ell^{\pm}\nu$, and the
irreducible process $\ee\ra\ZZ\ra\nn\qq$. Further
contributions are from events with particles escaping detection along the
beam-pipe, for example from $\Z$ boson decays accompanied by initial-state
photons or the untagged two-photon process $\ee\ra\ee\qq$. 
The following preselection is applied:

\begin{enumerate}
\item To reject non-hadronic events, at least $7$ charged particle
tracks are required. At least $20\%$ of all tracks must be of good 
quality (a minimum number of hits are required along the track, see 
Ref.~\cite{higgsold}); this is to reject badly measured events, mainly 
two-photon processes and beam-wall interactions. The total transverse momentum $p_{t}$
and the visible mass $m_{\mathrm{vis}}$ must satisfy $5\times p_{t}+m_{\mathrm{vis}}>\sqrts/2$,
and the visible energy $\mathrm{E}_{\mathrm{vis}}<0.8\sqrts$. The
energy measured in the forward detector components~\cite{detector}, 
which cover small polar angles, must be $<2\GeV$ in the forward calorimeters, $<5\GeV$ in the
gamma catcher and $<5\GeV$ in the silicon-tungsten calorimeter
\cite{opalSMHiggs172}. The overall energy observed in the region
$|\cos\theta|>0.9$, where $\theta$ denotes the polar angle with
respect to the electron beam, must not exceed $20\%$ of $\mathrm{E}_{\mathrm{vis}}$. 
\item The missing momentum vector has to point to sensitive parts of the
detector, $|\cos\theta_{\mathrm{miss}}|<0.95$, and the visible momentum
must not have a large component along the beam axis, $|p_{\mathrm{vis}}^{z}|<\sqrts/5$.
\item The tracks and clusters in each event are forced into two jets using
the Durham algorithm. Events with partially contained
jets  are rejected by the requirement $|\cos\theta_{\mathrm{jet}}|<0.95$
imposed on each jet.
\item $\Zgs\ra\qq$\ events are suppressed by requesting 
$\phi_{\mathrm{acop}}>5^{\circ}$
where the acoplanarity angle $\phi_{\mathrm{acop}}$ is the deviation of the angle
between the two jets in the plane perpendicular to the beam axis from $180^{\circ}$.
\item The missing mass, $m_{\mathrm{miss}}$, must be consistent with the Z boson
mass: $50\Gcs < m_{\mathrm{miss}} < 130\Gcs$.
\item Identified semi-leptonic \WW decays with energetic,
isolated~\cite{opalSMHiggs172} leptons are discarded.
\end{enumerate}
The signal likelihood function is constructed from 5 variables: 
(1)~$m_{\mathrm{miss}}$; (2)~$|\cos\theta_{\mathrm{miss}}|$;
(3)~$\textrm{max}|\cos\theta_{\mathrm{jet}}|$~i.e. the polar angle
of  the jet closest to the beam axis; 
(4)~the $\chi^{2}$-probability $\mathrm{P}(1\mathrm{C})$
of a one-constraint (1C) kinematic fit which imposes energy
and momentum conservation and constrains the missing mass to the 
Z boson mass;
(5)~the angle between the missing momentum and the jet with the higher 
energy: $\cos\theta_{\mathrm{j}-\mathrm{miss}}$. The distributions of
these discriminating variables are shown in Figure~\ref{figure5} 
and the likelihood distribution
in Figure~\ref{figure3}~(c). Events with a likelihood
larger than $0.4$ are selected. The Higgs boson mass is reconstructed
using the momenta provided by the 1C kinematic fit.

The number of events passing the likelihood selection is $123$  
(see Table 1) while $133\pm 11$
events are expected from SM background processes. 
The most important systematic uncertainties~\cite{opal183}
are from the modelling of the likelihood input
variables and from the lepton isolation
criteria. The signal efficiencies are affected by a total systematic error of
$2.9\%$. The Monte Carlo estimates of the signal efficiencies were reduced by 
$2.5\%$ to account for accidental vetoes due to accelerator-related backgrounds 
in the forward detectors. The reduction factor was determined from randomly triggered events.
\subsection{Searches in the electron and muon channels}
The signal events in the muon and electron channels are expected to have
two energetic, oppositely charged, isolated leptons and two hadronic jets. The
dominant backgrounds are $\ee\ra\Zgs$ accompanied by initial state
radiation and four-fermion processes, mainly from $\WW$ and $\ZZ$ pairs.
The preselection is described in the following:
\begin{enumerate}
\item Events without hadronic jets are rejected by requiring
at least 6 charged particle tracks. The visible energy $E_{\mathrm{vis}}$
must be larger than $0.6\,\sqrts$ and the component of the total momentum
along the beam axis must satisfy $|p_{\mathrm{vis}}^{z}|<E_{\mathrm{vis}}-0.5\,\sqrts$.
This requirement reduces $\ee\ra\Zgs\gamma\ra\qq\gamma$ and two-photon
processes, $\ee\ra\ee\qq$, significantly. All remaining events are
forced into four jets using the Durham algorithm allowing isolated
leptons to form low-multiplicity jets. Events are considered further
if the jet resolution parameter $y_{34}$ is larger than $10^{-4}$. 
\item Two oppositely charged electron or muon candidates must be identified,
with energies larger than $30\,(20)\GeV$ for the higher- (lower-) energy
candidate. The energy of muon candidates is deduced from the momentum
measurement in the central tracking chamber; for electron candidates the 
energy measured in the
electromagnetic calorimeter is used. The algorithms
to identify muons and electrons are described in \cite{opalMuonID}
and \cite{opalSMHiggs172}, respectively. 
\item The remaining part of the event, after the two lepton candidates are
removed, is reconstructed as a two-jet event using the Durham algorithm. 
If the lepton candidates
are muons, a 4C kinematic fit requiring energy and momentum
conservation is performed to improve the energy and mass
resolution of the muon pair;
the $\chi^{2}$-probability of the fit must exceed $10^{-5}$. 
For both electron and
muon candidate events, the invariant mass of the lepton pair is required to
be larger than $40\Gcs$.
\end{enumerate}
The signal likelihood is constructed from five variables in the muon channel
and nine variables in the electron channel. Those in common are:
(1) $E_{\mathrm{vis}}/\sqrts$; (2) $\log_{10}y_{34}$; (3-4)
the measured transverse momenta of the two lepton candidates ordered
by energy and calculated with respect to the nearest jet axis,
used to discriminate against semileptonic charm or bottom decays; 
(5) the invariant mass of the lepton pair. For each candidate in
the electron channel, the additional variables are: (6-7) $(E/p-1)/\sigma$
for the two electron candidates,
where the momentum $p$ is measured in the central tracking detector,
the energy $E$ is measured using the calorimeter and $\sigma$ denotes
the total error in $E/p$; (8-9) the normalised ionisation energy losses
in the central tracking chamber gas~\cite{opal183},
for the two electron candidates.
The event is selected if in the electron case the likelihood is
larger than 0.3 or in the muon case larger than 0.65.
Figures~\ref{figure3}~(d) and (e) 
show the distribution of the two likelihood functions. 
The mass recoiling against the lepton pair is taken as the reconstructed
Higgs boson mass.

The number of events passing the likelihood selection is $23$ in the electron
channel and  $16$ in the muon channel
(see Table 1) while the corresponding background expectations 
are $16.6\pm 5.1$
and $15.0\pm 2.9$ events.
Systematic uncertainties~\cite{opal183} arise mainly from 
the fragmentation
process, determined from a comparison of HERWIG and JETSET, and from 
different four-fermion rate predictions, 
given by grc4f, KORALW and EXCALIBUR.
The signal efficiencies have total systematic 
errors of less than 2\% .
\subsection{Search in the tau channel}
Signal events are expected to be composed of
two hadronic jets from the Higgs boson decay, and two tau leptons
from the $\Z$ decay. For each of the tau leptons, the decays into one or three 
charged particle tracks (``prongs") are considered, possibly accompanied
by calorimeter clusters from neutral particles. Important
sources of background are the processes $\ee\ra\ZZ^{(*)}\ra\qq\lpm$, 
$\ee\ra\WW\ra\qq\ell^{\pm}\nu$
and $\ee\ra\qq(\gamma)$. The following preselection is applied:
\begin{enumerate}
\item Events must be identified as multihadronic final states \cite{hadronic}.
The visible energy has to exceed $0.3\sqrts$. In order to reject
events in which particles escape detection close to the beam direction,
the missing momentum vector is required to point to sensitive detector
regions: $|\cos\theta_{\mathrm{miss}}|<0.95$. The scalar sum of the
transverse momenta of all measured particles has to be larger than
$45\Gc$.
\item Two isolated tau lepton candidates, each with a momentum between $15\Gc$
and $60\Gc$, are required. These are identified with artificial neural
networks (ANN) as described in~\cite{opal183}. Separate networks
are developed for 1-prong and 3-prong decays. From the ANN output, the probability
that a candidate is a real tau lepton is derived. The probabilities
$P_{1,2}$ of the two tau candidates are combined to a two-tau-likelihood:
${\cal L}_{\tau\tau}=P_{1}P_{2}/(P_{1}P_{2}+(1-P_{1})(1-P_{2}))$,
which must exceed $0.1$. If several tau pairs exist in the event, the pair with
the largest ${\cal L}_{\tau\tau}$ is chosen.
\item After removing the two tau candidates, the rest 
of the event is
grouped into two jets using the Durham algorithm. A kinematic
fit (2C) is applied to the momenta of the two tau candidates and the
two reconstructed jets, imposing energy and momentum conservation.
The directions of the tau candidates are approximated by the visible
momenta of their decay products; their energies are free parameters
in the fit.
The $\chi^{2}$-probability of the fit must be larger than $10^{-5}$.
\item If both tau decays are classified as 1-prong decays, the momentum
sum of both charged tracks must be less than $80\Gc$; this is 
to reduce backgrounds
from $\ee\ra\ZZ\ra\qq\mumu$\ and $\qq\ee$.
\end{enumerate}
The signal likelihood is constructed
using nine variables: (1) the visible energy; (2) $|\cos\theta_{\mathrm{miss}}|$;
(3) $y_{34}$ obtained after reconstructing  the event, including the tracks 
and clusters of the tau candidates, into four jets (the Durham algorithm is
used); (4-5) the angles between
each of the two tau candidates and the nearest jet; (6) the energy
of the most energetic muon or electron, if any; (7) the $\chi^{2}$-probability
of a 3C kinematic fit, which in comparison to the 2C fit, in addition constrains
the invariant mass of the two tau candidates to the $\Z$
boson mass; (8) the two-tau likelihood ${\cal L}_{\tau\tau}$; (9) the impact 
parameter joint
probability of the tau candidate tracks calculated as in
Ref.~\cite{ALEPHImpactJointProb}. 
The resulting likelihood distribution is shown in Figure~\ref{figure3}~(f).
Events with a likelihood larger than $0.8$ are accepted. The invariant
mass of the two jets, resulting from the 3C-fit, is taken as the reconstructed
Higgs boson mass.

Three events pass the likelihood cut (see Table 1) compared to $8.8\pm 1.5$
events expected from background. The systematic errors 
are determined as described in~\cite{opal183}. The largest uncertainty
arises from the purity of the tau lepton selection. 
The signal efficiencies are affected by a total systematic 
error of $15-17\%$.
\section{Results}
All search channels combined, $604$ candidates are selected, while  
$588\pm56$ are expected from background processes (these numbers apply
for a test-mass of $100\Gcs$\ in the four-jet channel).
Figure~\ref{figure4}~(b) shows the distribution of the reconstructed Higgs 
boson mass for the candidates selected in the missing energy,
electron, muon and tau channels, and for the corresponding expected 
backgrounds, added together.

No significant excess is observed in any of the search channels 
over the expected background from SM processes. 
In the following, an upper limit is set 
on the product of the cross-section $\sigma_{\hZ}$ of the 
Higgsstrahlung process and the hadronic branching ratio $\mathrm{Br}(\h\ra\mathrm{hadrons})$
of the Higgs boson. For this purpose, these search results
are combined with previous OPAL results, obtained at 
$\sqrts=91\GeV$ in the missing energy, electron and muon 
channels~\cite{opalSMHiggsLEP1}, and at $\sqrts=189\GeV$
in all channels~\cite{opal2HDM189}.

The limits are obtained by using a weighted event counting method
(see Section~5 of \cite{mssmpaper172}). The systematic errors
are incorporated following Ref.~\cite{cousins}. 
The four analyses are split into several sub-channels denoted by $i$ 
depending on the value of the discriminating variable. Initially the 
weight, $\omega_{i}=s_{i}\ /\ (s_{i}+b_{i})$, is assigned to each channel,
where $s_{i}$ and $b_{i}$ are the signal and background expectation, respectively.
These weights are a good choice if the systematic errors are negligible
or do not differ among the channels. To obtain weights which are optimal for the
general case, the weights are varied until the
the variance of the distribution of $\sum_i \omega_{i}\ (s_i + b_i)$ becomes minimal while
keeping $\sum_{i} \omega_{i} s_{i}$ constant.
The variance is given by
$$
\sigma^2=\sum_{i} \omega^{2}_{i}\ (s_i+b_i)+\sum_{l}\ \left\{\sum_i \omega_i\ (s_{i} \sigma^{s}_{li}+b_{i} \sigma^{b}_{li})\right\}^2,
$$
where $l$ denotes the independent  error sources which give rise to the systematic 
errors on the signal, $s_i \sigma^{s}_{l_i}$, and on the background, $b_i \sigma^{b}_{l_i}$, of the
various channels.
A unique solution, which minimises the variance, is guaranteed requiring $\omega_i\geq 0$.  
To fulfil this constraint, an unambiguously defined set of sub-channels has to be dropped. 

\begin{figure}[htb]
\begin{center}
\epsfig{figure=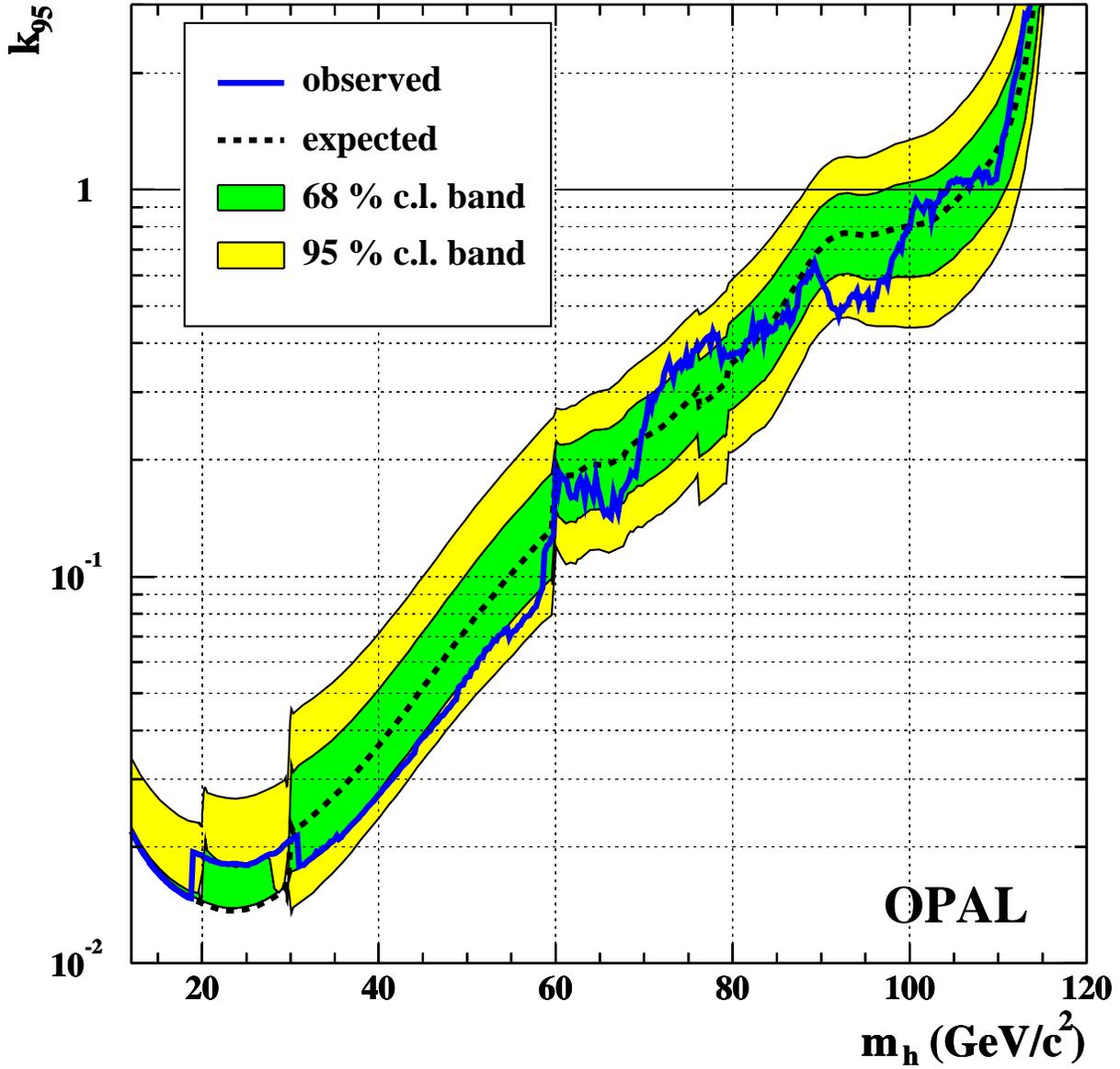,width=\textwidth}
\caption[]{\small The $95\%$ confidence level upper bound on the product $k$ of the Higgsstrahlung
cross-section and the hadronic decay branching ratio of the Higgs boson,
divided by the Higgsstrahlung cross-section in the SM.
The thick solid line shows the observed limit. The limit expected on average, in a
large number of simulated experiments,
in the absence of a Higgsstrahlung signal is indicated by the dashed line
while the dark- and light-shaded areas show the $68\%$ and $95\%$ probability bands
around the average.}
\label{figure6}
\end{center}
\end{figure}

A test-mass dependent $95\%$ confidence level upper bound $k_{95}$ is calculated for the quantity  
$$
k(\mh) = \frac{\sigma_{\hZ}(\mh)\times\mathrm{Br}(\h\to\mathrm{hadrons})}
{\sigma_{\mathrm{HZ}}^{\mathrm{SM}}(\mh)}
$$
where $\sigma_{\mathrm{HZ}}^{\mathrm{SM}}$ is the predicted SM cross-section 
for the Higgsstrahlung process. This bound is shown in Figure~\ref{figure6}. 
In calculating this limit,
the four-jet and tau channels were considered only for masses 
above $60\Gcs$\ 
while the other channels contributed from $30\Gcs$\ upwards. 
Between $12$ and $30\Gcs$,
only the data taken in the vicinity of $\sqrts=91\GeV$ are used~\cite{opalSMHiggsLEP1}.
The region below $12\Gcs$\ is covered by a decay mode independent
Higgs boson search conducted by OPAL~\cite{OPALDecayModeIndep}.

Limits on the cross-section $\sigma$
for arbitrary $\mathrm{Br}(\h\ra\mathrm{hadrons})$
or for arbitrary $\hZ$ coupling strength $g_{\hZ}$
can be derived using the expression
$$
\sigma_{95} = k_{95}\times\sigma_{\mathrm{HZ}}^{\mathrm{SM}}\times\mathrm{Br}
(\h\to{\mathrm{hadrons}})\times(g_{\hZ}/g_{\mathrm{HZ}}
^{\mathrm{SM}})^{2},
$$
provided that the effective hZ coupling has the SM Lorentz structure. 

Assuming the hZ coupling predicted by the SM, a Higgs boson
decaying only into hadronic final states ($k_{95}(\mh)=1$) is excluded for masses up to $104\Gcs$.
For a Higgs boson also having the decay properties predicted by the SM, 
this limit is at $100\Gcs$.
\section{Summary\label{sec:summary}}
A search has been performed for a hypothetical neutral scalar Higgs boson 
which is produced in Higgsstrahlung and which decays to hadrons of arbitrary 
flavour. The search 
is based on data collected by the OPAL experiment in \ee\ collisions at
centre-of-mass
energies between $192$ and $209\GeV$. The results have been combined with
earlier OPAL searches conducted at $\sqrt{s}\approx 91\GeV$ and 
$\sqrts = 189\GeV$.
No significant excess has been observed over the background 
expected from Standard Model processes. 
A mass-dependent upper bound is set, at the $95\%$ confidence level,
on the product of the Higgsstrahlung cross-section and the 
hadronic branching ratio of the Higgs boson. 
For a Higgs boson which couples to the $\Z$ boson with Standard Model strength and
which decays exclusively into hadronic final states, a flavour independent 
lower bound of $104\Gcs$\ is obtained on the mass.

\section*{Acknowledgements}
We particularly wish to thank the SL Division for the efficient operation
of the LEP accelerator at all energies
 and for their close cooperation with
our experimental group.  In addition to the support staff at our own
institutions we are pleased to acknowledge the  \\
Department of Energy, USA, \\
National Science Foundation, USA, \\
Particle Physics and Astronomy Research Council, UK, \\
Natural Sciences and Engineering Research Council, Canada, \\
Israel Science Foundation, administered by the Israel
Academy of Science and Humanities, \\
Benoziyo Center for High Energy Physics,\\
Japanese Ministry of Education, Culture, Sports, Science and
Technology (MEXT) and a grant under the MEXT International
Science Research Program,\\
Japanese Society for the Promotion of Science (JSPS),\\
German Israeli Bi-national Science Foundation (GIF), \\
Bundesministerium f\"ur Bildung und Forschung, Germany, \\
National Research Council of Canada, \\
Hungarian Foundation for Scientific Research, OTKA T-038240, 
and T-042864,\\
The NWO/NATO Fund for Scientific Research, the Netherlands.\\

%
%
%
%
%
%
%
%
%
%
%
%
%

\end{document}